\documentclass[aps,prl,twocolumn,superscriptaddress]{revtex4}

\usepackage[dvips]{graphicx}
\usepackage{amssymb,amsfonts,amsmath}
\usepackage{bbm, bm}

\begin{document}
\title{Theoretical analysis of the role of chromatin interactions in long-range action of enhancers and insulators}

\author{Swagatam Mukhopadhyay}\affiliation{Cold Spring Harbor Laboratory, Cold Spring Harbor, NY USA}
\author{Paul D. Schedl}\affiliation{Department of Molecular Biology, Princeton University, Princeton, NJ USA}
\author{Vasily M. Studitsky}\affiliation{Department of Pharmacology, UMDNJ-Robert Wood Johnson Medical School, Piscataway, NJ, USA }

\author{Anirvan M. Sengupta}\affiliation{Department of Physics and Astronomy and BioMaPS, Rutgers University, Piscataway, NJ USA }

\date{\today}

\begin{abstract}
Long-distance regulatory interactions between enhancers and their target genes are commonplace in higher eukaryotes. Interposed boundaries or insulators are able to block these long distance regulatory interactions. The mechanistic basis for insulator activity and how it relates to enhancer action-at-a-distance remains unclear. Here we explore the idea that topological loops could simultaneously account for regulatory interactions of distal enhancers and the insulating activity of boundary elements.  We show that while loop formation is not in itself sufficient to explain action at a distance, incorporating transient non-specific and moderate attractive interactions between the chromatin fibers strongly enhances long-distance regulatory interactions and is sufficient to generate a euchromatin-like state. Under these same conditions, the subdivision of the loop into two topologically independent loops by insulators inhibits inter-domain interactions. The underlying cause of this effect is a suppression of crossings in the contact map at intermediate distances. Thus our model simultaneously accounts for regulatory interactions at a distance and the insulator activity of boundary elements.  This unified model of the regulatory roles of chromatin loops makes several testable predictions that could be confronted with \emph{in vitro} experiments, as well as genomic chromatin conformation capture and fluorescent microscopic approaches.
\end{abstract}
\maketitle

Unlike most known cases of transcriptional regulation in prokaryotes and lower eukaryotes, metazoan genes are often regulated by enhancers placed tens to hundreds of kilobases away from the promoter~\cite{Blackwood1998, Dorsett1999, Nobrega2003, West2005}. Facilitating mechanisms are necessary for such long-range enhancer action, as we shall explain below. Widespread distant regulation also requires additional mechanisms to ensure specificity. Enhancer-blocking DNA sequences, known as boundaries or insulators, define chromatin domains within which enhancer action is limited ~\cite{Bondarenko2003, Bushey2008, Capelson2004, Corces1995, Gaszner2006, Valenzuela2006, Wallace2007}. While it is known that insulator elements bind to particular proteins~\cite{Cuddapah2009} how these protein complexes manage to block enhancer action across domains remains controversial. 

Several different models for long-range enhancer-promoter communication have been proposed, for review see~\cite{Bondarenko2003, Valenzuela2006}. One model hypothesizes a tracking mechanism that involves the processive movement of regulatory machines launched from the enhancer towards the promoter. Another model hypothesizes that transcriptional up-regulation requires direct physical contact between proteins assembled at the enhancer and the transcriptional apparatus at promoter. This process necessarily leads to looping out the intervening chromatin.  Looping model has received significant support in the context of the control of the beta-globin locus by the LCR~\cite{Carter2002, Tolhuis2002}. 
For each of these models of enhancer-promoter communication, one needs a corresponding mechanism of action for insulators~\cite{Gaszner2006, Geyer2002, Majumder2003, Valenzuela2006}. For the tracking model, insulators are assumed to work as barriers blocking the movement of the regulatory machine.  In the looping model, insulators function by decoying promoters or other acting as sinks or traps for enhancer~\cite{Geyer1997}. 
 
Yet another model for enhancer action is based on the idea that insulators subdivide the genome into topologically independent loops or domains~\cite{Gerasimova2001}.  In this model, enhancer action at distance requires a mechanism that promotes intra-loop enhancer-promoter contacts, while insulation requires that inter-loop contacts be disfavored.  The topological loop model does seem to explain experiments that aim to contrast conjectured mechanisms of insulation~\cite{Gohl2010, Gohl2008, Savitskaya2006}; however, scant attention has been paid to the question of whether the topological loop model is plausible from a physical point of view. We redress this critical gap in our understanding of long-range gene regulation. Specifically, we resolve the following puzzles within the context of looping models--- 
\begin{itemize}
\item{What are the ingredients necessary in a physical model of chromatin capable of producing efficient long-range enhancer-promoter communication?}
\item{What are the signatures of such physical features on observable conformations of chromatin? What conformations are favored?}
\item{What are the consequences of favored chromatin conformations on insulation by insulators arranging chromatin into topological loop domains?} 
\end{itemize} 
Surprisingly, we discover that the same model that explains experimentally observed efficiency of long-range enhancer action is, paradoxically, capable of explaining the efficiency of insulator action; no added ingredients are needed. 

\section{Results}

Our intention is to understand dynamics of large domains of chromatin ranging in size from tens to hundreds of kilobases of DNA.  \emph{Ab initio} molecular modeling of such large systems is hindered both by computational limitations and our lack of detailed knowledge of chromatin composition and structure. Moreover, keeping in mind that robust predictions can often be extracted from coarse-grained models, we use such an approach.

\subsection{Long-range enhancer action}
\label{sec:long-range}
\begin{figure}
\begin{center}
\centerline{\includegraphics[width=0.6\textwidth]{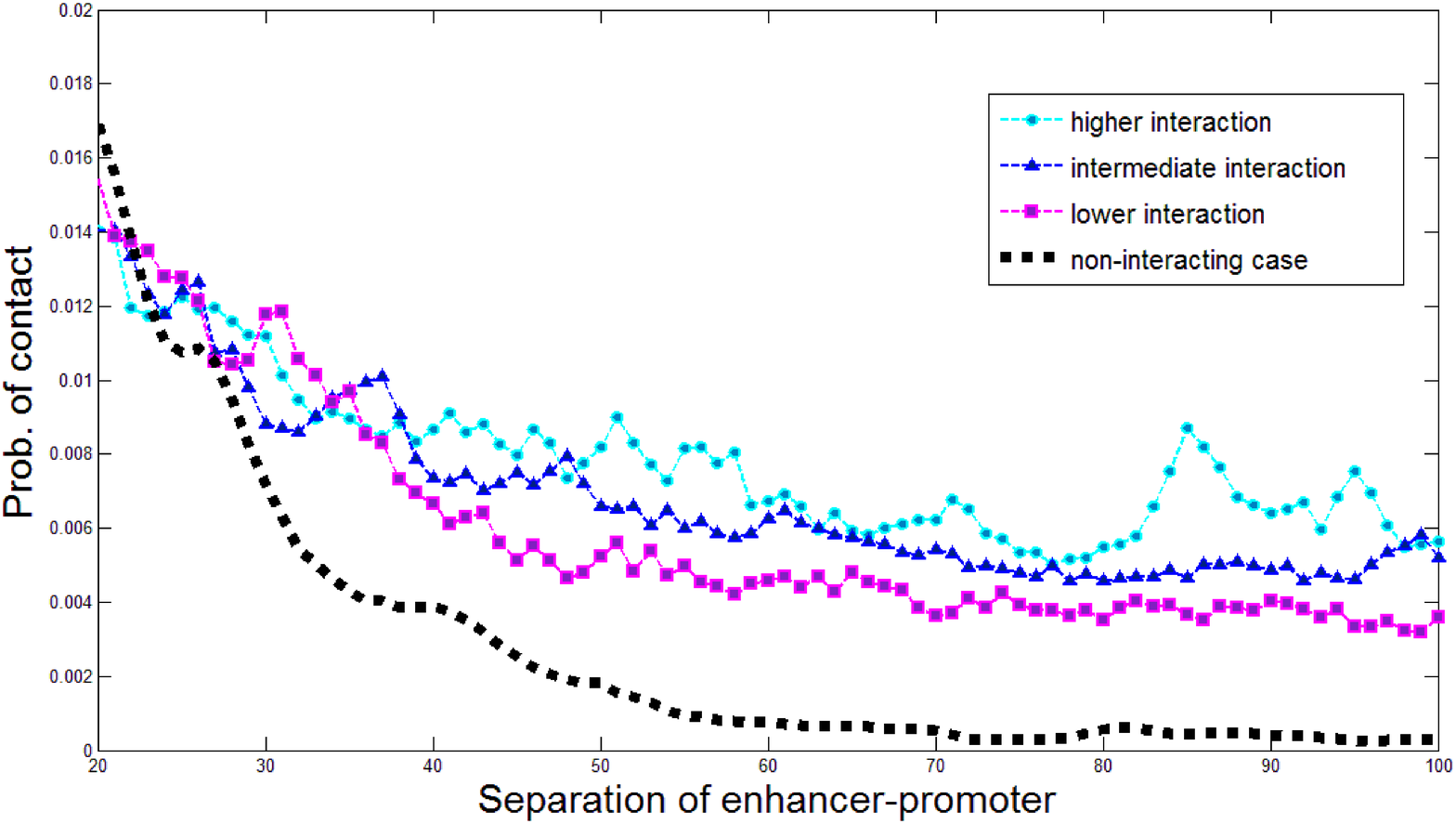}} 
\caption{Contact probability for fixed availability and different  attraction, zoomed in on the tail of the distribution, against enhancer-promoter separation. The typical contact probability is peaked around a separation determined by the choice of persistence length, and is excluded from this plot. As expected, the tail of the distribution falls off as a power law for the non-interacting case,  but has a much slower fall-off for our interaction model. We present data for three different values of effective interaction and show that higher interaction leads to better long-range communication.}
\label{fig:Figure1}
\end{center} 
\end{figure}

Distance dependence of contact probability between two points on a semi-flexible polymer has been studied under many contexts~\cite{Shimada1984}. The contact probability is the highest for a separation of the order the persistence length (which is known to be a few nucleosomes in the case of chromatin~\cite{Cui2000,Dekker2002} and falls off rapidly as a power law for longer distances, see~Figure~\ref{fig:Figure1}. At separations of tens to hundreds of kb the contact probability should have fallen from its peak value to few orders of magnitude below, contrary to observations. Quite obviously, the semi-flexible polymer properties of chromatin alone is incapable of reproducing efficient long-range enhancer-promoter contacts. A plausible argument explaining the phenomenon is very strong enhancer-promoter interaction; the enhancer stays in contact with the promoter for a long time once (an otherwise improbable) contact is established. Had there been only one promoter, this mechanism could perhaps explain the enhanced level of \emph{average} gene expression at long ranges. However, in this scenario, if the same enhancer has multiple competing promoters at different distances, the proximal promoters would always be favored for contact, contrary to experimental observations~\cite{Tsytsykova2007}. Moreover, multiple enhancers activating the same promoter is difficult to arrange within this scenario~\cite{Deschenes2007}.

Consequently, besides the semi-flexible polymer features, we need to recognize some other feature of chromatin that might favor long distance contacts. We argue that the new ingredient is chromatin-chromatin attractive interactions. There are many potential sources of such an interaction. One possibility that has been explored experimentally is histone-tail mediated inter-nucleosome interaction~\cite{Angelov2001, Arya2009, Arya2006, Moore1997, Zheng2003}. Another mechanism involving nucleosomes would be electrostatic interactions between histone cores~\cite{Chodaparambil2007, Hizume2005, Luger1997, Zhou2007}. Alternatively, the many different DNA/chromatin binding proteins, known for promoting chromatin association and condensation, might mediate `nucleosome-nucleosome' interactions. Examples of such proteins include the linker histones H1 and H5~\cite{Bustin2005, Grigoryev2009}, HMG proteins \cite{Phair2004, Rochman2009} and HP1~\cite{Nielsen2001}.  However, our conclusions do not depend upon the precise cause or causes of this postulated attractive chromatin-chromatin interaction. Rather we simply ask whether incorporating some type of transient and weak chromatin-chromatin attractions into our model is sufficient to generate efficient communication between distant enhancers and promoters. 		
	
What sort of attractive interactions should one consider? It is well known that polymer interactions with a static attraction potential (even if the potential is short-range) can lead to polymer agglomeration for potentials with strength of the order of thermal fluctuation energy $k_BT$ (which is about a tenth of a typical hydrogen bond). This phenomenon is known as the \emph{coil-globule transition}. Many authors~\cite{Bystricky2004, Langowski2007, Nakai2005, Schiessel2003, Widom1998} have drawn analogies with the coil-globule transition and its relation to compaction of chromatin, including formation of \emph{heterochromatin}. However, we are interested in the role of transient nucleosome-nucleosome interaction in \emph{euchromatin}. Euchromatin conformations are known to be extended~\cite{Hahnfeldt1993, Engh1992}. Motivated by the idea that an additional degree of freedom (tail configuration or protein binding) is the mediator, we set up our effective model of bead-to-bead interaction as follows:
Instead of a uniformly attractive interaction, we introduce two discrete states for each nucleosome (\emph{inert} and \emph{active}) that it switches between stochastically. Only \emph{active} nucleosomes can form an attractive bond within a short interaction range, see details in Materials and Methods. 
	
Figure~\ref{fig:Figure1} shows that when transient nucleosome-nucleosome interactions are incorporated into the model, enhancer promoter interactions do not fall off rapidly with enhancer-promoter separation along the chromatin. Instead, the probability of enhancer-promoter contacts stays relatively constant over a wide range of distances, as is also seen in experimental studies~\cite{Polikanov2008}. Moreover, with incremental increases in interactions, the probability of long-range contact is increased. We have verified this key result for a wide range of model parameters including polymer stiffness (data not shown for this parameter). This nearly flat tail of the contact probability is a robust feature of our model for intermediate separations of enhancer promoter pairs, and is in striking contrast to the corresponding behavior of either non-interacting or collapsed polymers, see Figure~\ref{fig:Figure1}. Therefore, we identify novel aspects of chromatin polymer that is capable of explaining long-range action---
\begin{itemize} 
\item{Nucleosomes are only weakly and transiently mutually attractive.}
\item{The pair-wise and higher order nucleosome interactions are not identical.}
\end{itemize} 
For simplicity, we limit our model to attractive pair-wise nucleosome interactions and repulsive higher order interactions. This construction averts polymer collapse, see further discussion in Methods. 

\subsection{Analysis of enhancer-promoter interaction}

\begin{figure}
\begin{center}
\centerline{\includegraphics[width=0.7\textwidth]{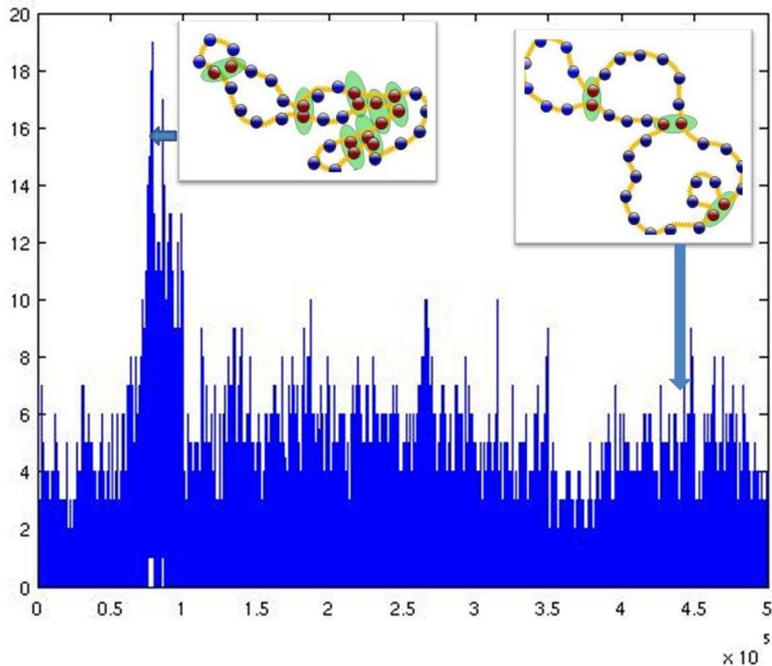}} 
\caption{Timeline plot (MC steps) of number of transient bonds formed in a window of a typical run. The red beads are the active beads that has formed transient bonds (green shadow) and the blue beads are inert. Portion of the 200 bead configuration shown. The highly contacted configurations are typically more compact.}
\label{fig:Figure2}
\end{center} 
\end{figure}

\begin{figure}
\begin{center}
\centerline{\includegraphics[width=0.7\textwidth]{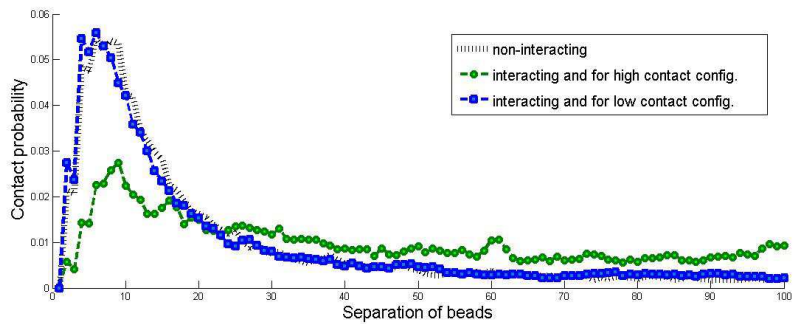}} 
\caption{Probability of contact for two different contact sizes, low and high.}
\label{fig:Figure3}
\end{center} 
\end{figure}

	We have shown that it is possible to make efficient long-range contacts in our model of euchromatin without having to undergo a collapse transition. This result is attractive for the following reason. Experiments measuring physical distance as a function of genomic distance are consistent with the euchromatin configuration statistics being roughly given by Gaussian random walks~\cite{Bystricky2004, Goetze2007, Hahnfeldt1993, Engh1992}. This behavior is puzzling because contacts happen efficiently on euchromatin in spite of the apparent Gaussian statistics. We resolve this puzzle by looking at the time dependence of number of contacts in our model, see Figure~\ref{fig:Figure2}. We observe that the configurations with many contacts are sporadic and relatively rare. The polymer goes through occasional compact configurations that are responsible for the enhancement of long-range contact probability; however, consistent with experimental measurements, the typical configuration statistics is roughly Gaussian. We prove this claim by separating out the contact probabilities for configurations with low and high number of total instantaneous contacts, see Figure~\ref{fig:Figure3}. The key observation is that long-range contact probability is overwhelmingly contributed by high-contact configurations. 
	
\begin{figure}
\begin{center}
\centerline{\includegraphics[width=0.7\textwidth]{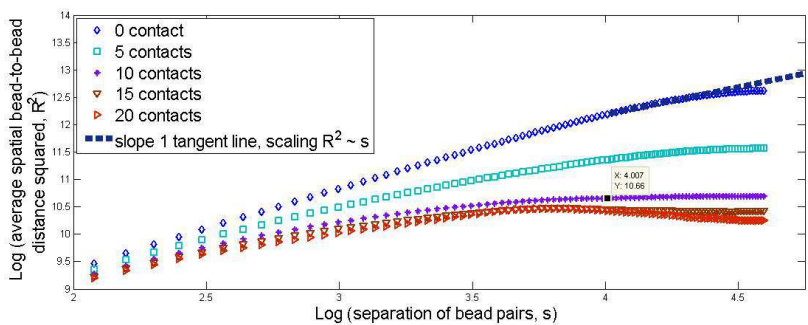}} 
\caption{Log-log plot of RMS distance of bead pairs against separation for configurations selected by their number of contacts. The zero contact configurations are nearly Gaussian; deviation from the random walk scaling at short separations is owing to persistence of the polymer and at long separations owing to the finite size effect of the ring configuration studied here. For high number of contacts we see significant deviation from the random walk behavior; the polymer is more compact. However, these configurations are sporadic and rarer.}
\label{fig:Figure4}
\end{center} 
\end{figure}

We unravel the Gaussian nature of configurations with low number of contacts more directly by measuring the root-mean-square (RMS) bead-to-bead distance as a function of bead separation for configurations selected by their number of contacts. For a Gaussian polymer the RMS distance should scale as the square root of separation, for separations beyond persistence length. Figure~\ref{fig:Figure4} presents the log-log plot of the RMS distance against separation. It shows that for low number of contacts the behavior of the RMS distance is indeed roughly Gaussian, and for highly contacted configurations significant deviation from Gaussian behavior is observed. We have carried out similar analysis for the pair-correlation function (data not shown). Because the high-contact configurations are rarer, the crucial conclusion form this analysis is that a typical configuration will exhibit Gaussian statistics, in agreement with experimental observations. 

It turns out that the structure and sporadic behavior of these high contact configurations is of crucial importance in understanding how insulators could potentially block enhancer action within this model. We emphasize that our model for long-range enhancer action is not augmented in any manner for our study of insulator action described below. 

\subsection{Insulator action} 

In the topological loop model the key function of insulators is to define the two end-points of each looped domain. How they do so is a matter of conjecture. One idea is that insulators for loops by associating with the nuclear matrix or insulator bodies.  Another idea is that insulators generate loops by pairing directly with each other. 
	
\begin{figure}
\begin{center}
\centerline{\includegraphics[width=0.5\textwidth]{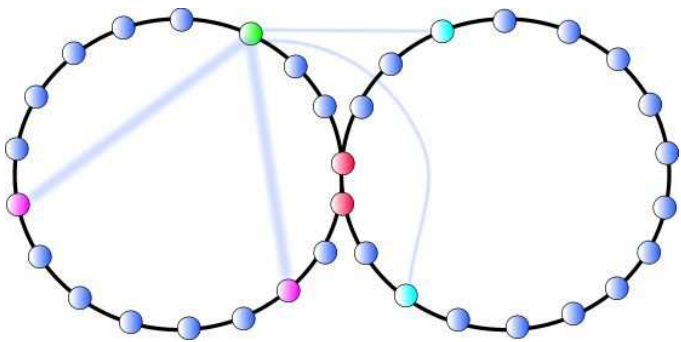}} 
\caption{Set up for studying insulator action; red beads are insulators, magenta beads are equi-distant intra-domain promoters and cyan beads are equi-distant inter-domain promoters.}
\label{fig:Figure5}
\end{center} 
\end{figure}

In order to study the simplest scenario for insulator dependent loop formation, we construct two equal domains by pinching a ring configuration of chromatin in the middle by a permanent bond between insulators.  This would simulate insulator pairing, and of the possible mechanism for insulator dependent loop formation, \emph{it is most exigent on the chromatin model}. It is not at all obvious why in such a set-up, inter-domain contacts could be suppressed in comparison to intra-domain contacts (Figure~\ref{fig:Figure5}). 

\begin{figure}
\begin{center}
\centerline{\includegraphics[width=0.7\textwidth]{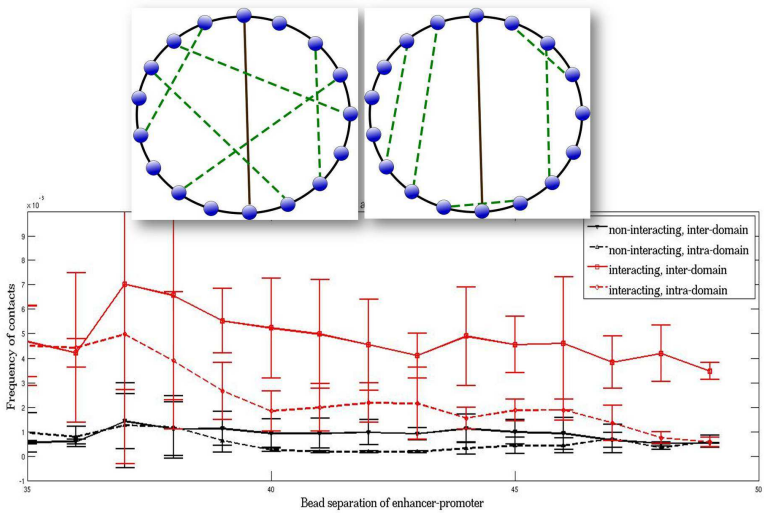}} 
\caption{Efficiency of insulation for interacting polymer model.The upper panel is a sketch of a typical contact map for the same number of contacts for random distribution of contacts (on the left) and for our simulation of interacting nucleosomes (on the right). The solid line is the insulator contact (permanent) and dashed lines represent transient contacts between nucleosomes. Interaction reduces long-range crossing of contacts as quantified in the lower panel graph by comparing inter- and intra-domain contacts.}
\label{fig:Figure6}
\end{center} 
\end{figure}

We simulate the same model used for studying enhancer-promoter communication, and vary the interaction strength in the same fashion. We query the probability of inter-domain vs. intra-domain contact by measuring the contact frequency between a bead (enhancer) and its four equidistant neighbors (promoters), two of which are inter-and two are intra-domain, see Figure~\ref{fig:Figure5}. We measure these probabilities for various separations of these beads. 	
Two findings are significant, see Figure~\ref{fig:Figure6}.  First, as was the case for the simple ring, weak and transient nucleosome-nucleosome attraction promotes inter-domain "enhancer-promoter" interactions, and the frequency of these long distance "regulatory" interactions increases with increasing attraction. Second, while transient nucleosome-nucleosome attraction supports distant intra-domain enhancer-promoter interactions, it also suppresses enhancer-promoter interactions between elements in different loop domains, see Figure~\ref{fig:Figure6}. As with intra-domain interactions, the suppression of inter-domain interactions is correlated with the strength of the transient nucleosome-nucleosome attraction. Compared to the non-interacting case, the suppression factor can be quite large (Figure~\ref{fig:Figure6}), and in addition to the strength of nucleosome attraction it also depends on how deep within a domain the enhancer is. This is in close agreement with the quality of insulation observed as a function of distance from the insulator in experiments~\cite{Savitskaya2006, Scott1999}. In this context, it is worth noting that the \emph{barrier} and the \emph{decoy insulator models} of insulation cannot explain why the efficiency of insulation depends on the distance of the enhancer and promoter pair from the insulator pair bracketing it. In the topological loop model our data explains the phenomena as follows.  When the enhancer and promoter are proximal to paired insulators (that create the loop domain), insulation is not as efficient because only further away from a certain proximity to insulators is insulation facilitated in our model. The origin of this length scale and the dependence of the efficiency of insulation on the interaction set-up are analyzed in the next section.  

\subsection{Understanding insulator action}
One seemingly paradoxical feature of our model is that weak transient nucleosome-nucleosome attraction on the one hand is able to promote long distance intra-domain enhancer-promoter communication and on the other hand is able to suppress enhancer-promoter communication across domains. In order to resolve this paradox, we analyzed the topological structure of the contact map.  For any conformation of our model polymer, one can map the instantaneous contacts, formed during the sporadic events, into chords on a discrete circle where the end points of the chords are the beads in contact. For example, the permanent contact introduced between insulators is a diameter in this mapping. Inter-domain/intra-domain interactions correspond to chords that cross/do not cross this diameter. We analyze the crossing of these chords in our model simulation for enhancer action to elucidate insulation action as follows.

\begin{figure}
\begin{center}
\centerline{\includegraphics[width=0.7\textwidth]{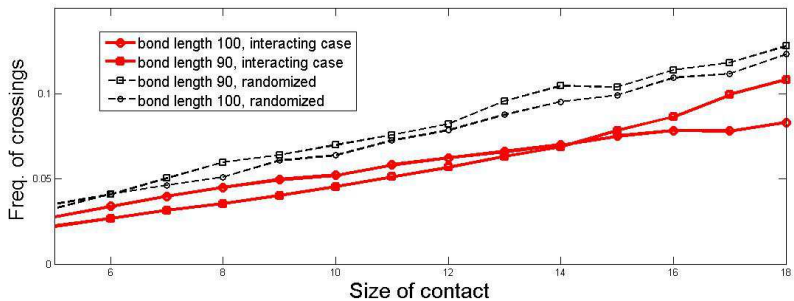}} 
\caption{Frequency of crossing of chords (discussion of mapping of contacts to chords on a circle is in the text) with a chosen chord of fixed length, against contact sizes. Data shown for two lengths of chosen chord. The 100 length chord is a diameter in our 200 bead simulation, and therefore corresponds to the insulator setup. For comparison we have randomized the chords  and computed the frequency of crossing. No insulation is expected for random contacts between any two beads, therefore the randomized chords serves as a benchmark.}
\label{fig:Figure7}
\end{center} 
\end{figure}

We select conformations by their number of instantaneous contacts (\emph{contact size}). Because we are interested in long-range action, we only consider contacts between beads separated by more than a short-range cut-off. We pick this cutoff to be the separation beyond which the probability of contact for non-interacting polymer begins to depart significantly from that of the interacting polymer (Figure~\ref{fig:Figure3}).  For each contact size, we compute the statistics of number of crossings of all mapped chords with a single chord of fixed length. We perform this analysis for various lengths of the chosen chord.  As shown in Figure~\ref{fig:Figure7}, the central observation is that the frequency of crossing of a long-range contact is consistently lower in the interacting polymer, in comparison to when chords are randomly distributed on a circle. If the long-range contact linking the loop is permanent, as it is in the insulator simulation (Figure~\ref{fig:Figure5}, the observation immediately implies that inter-domain contact is suppressed (Figure~\ref{fig:Figure6}). For random chords on a circle we do not expect any such suppression, and hence this acts as a benchmark for efficiency of insulation (Figure~\ref{fig:Figure7}). Note that suppression of crossing is poor for short-range contacts, i.e., an \emph{interaction dependent length-scale} for efficient insulation emerges in this picture, as discussed earlier. 
	
The contact map statistics also implies that the high contact configurations are not just temporary appearance of a state of fully collapsed polymer. In polymer literature, pseudo-knots in attractive polymers above the collapse transition have been explored, motivated mostly by RNA folding~\cite{Kabakcioglu2004}. Our results are consistent with picture that a single contact usually does not form a pseudo-knot by crossing a large number of other contacts. The entropic cost of a heavily crossed contact map makes such contact patterns less likely. We have verified (data not shown) that increasing nucleosome-nucleosome attraction does reduce the overall number of crossings of all long-range contacts for a given contact size. However this reduction, along with insulation, is non-trivially dependent on the effective interaction strength given by the combination of \emph{attraction} and \emph{availability} (see Methods). The subtle relationship between geometry of the double loop configuration and insulation is controlled by the typical contact size, which in turn is controlled by the effective interaction strength. This interaction strength sets the length scale beyond which contact crossing is suppressed.  A detailed exploration of the putative reduction in pseudo-knots of the contact map (and ensuing insulation) is beyond the scope of this paper, and will be presented elsewhere.

\section{Discussion}

	We have presented an attractive and unifying theory of two central facets of gene regulation in eukaryotes: enhancer action at distance and enhancer (silencer) blocking activity of insulators. We show that the topological loop model accounts not only for long distance regulation by enhancers (silencers) but also for the ability of insulators to restrict the action of regulatory elements to the domain in which they reside.  It is possible to reconstruct both of these regulatory phenomena in the context of the topological loop model with only one key but plausible assumption, namely that there are weak attractive interactions between nucleosomes or other generic chromatin components. By incorporating this feature into the properties of the chromatin polymer, the topological loop model not only explains the experimentally observed slow decay of long-range enhancer-promoter communication, but also the ability of insulators to block inter-domain regulatory interactions.
	
One limitation of our model is that the chromatin fiber is simulated by a polymer whose structural properties approximate the nucleosomal beads-on-a-string configuration.  In fact, chromatin in the cell has an as yet not fully understood higher order structure that substantially condenses the chromatin fiber.  In the context of a looped domain, this compaction would bring distant points in a loop in much closer contact than would be the case for a beads-on-a-string polymer.  As a first attempt to simulate the effects of higher order structure on the regulatory properties conferred by subdividing chromatin into topologically independent loops, we examined the effects of a modest supercoiling of the chromatin polymer.  Significantly, we found that introducing just a low number of supercoils substantially augments long distance enhancer-promoter interactions within a looped domain, and leads to an even greater suppression of inter loop domain regulatory interactions than seen in the model presented here. 

\subsection{Proposal for experimental investigation} 

Our computational model not only expounds a possible mechanism for the hitherto unexplained efficacy of loop-domains in long-range gene regulation, it offers concrete and testable predictions. Briefly, they are---
\begin{itemize} 
\item{Though euchromatin conformations are \emph{typically} Gaussian polymer in nature, rare compact conformations appear sporadically, and these conformations predominantly contribute to long-range contact probability.}
\item{In the presence of multiple insulators, the specific loop domains formed by insulator pairing/clustering determine the suppression of inter-domain contacts and this suppression is dependent on the relative distances and positions of enhancers, promoters and insulators.} 
\end{itemize} 
Our model also offers a quantitative framework for investigating these predictions. Testing the first prediction is possible by direct visualization of chromatin conformation by \emph{in situ} techniques like high-resolution multicolor 3D-FISH~\cite{Cremer2008} or multicolor 3D-SIM~\cite{Schermelleh2008}. Both of these techniques have the advantage of high-resolution; fluorescent probes can be used to tag tens of kilobases on chromatin where the total number of tags is limited by the total number of colors observable. A suitable genomic locus of length $100Kb-1Mb$ can be investigated in such techniques. Briefly, our model would predict that in a gene-poor region of euchromatin the statistics (averaged over many nuclei) of the three dimensional organization of the multi-color probe-signals would be approximately Gaussian. However, rare individual conformations would be observed where the colors are distributed in a compact conformation inexplicable by Gaussian statistics. We propose studying gene-poor region because the inter-nucleosomal interactions that lead to such sporadic compaction is non-specific in nature. 

Testing the second prediction is possible by studying insulator-rich genomic loci using the same techniques.  Knowledge of the genomic location of insulators (like CTCF) and the pairing/clustering interactions of insulator elements would allow one to resolve the statistics of inter-domain and intra-domain contacts using fluorescent microscopy. Another technique that can be used to complement such a quest is Chromatin Conformation Capture (3C/5C)~\cite{Dekker2002, Dostie2006, Simonis2007}, which map the pair-wise contact frequency between multiple regions of the chromatin. It has the advantage of being able to simultaneously query contacts between any number of `points' on chromatin, but the disadvantage that only \emph{average} and \emph{pair-wise} contacts are observable. The 3D conformation of chromatin cannot be extracted. Nevertheless, coupled with fluorescent microscopic techniques, 3C/5C can provide crucial information on the probable insulator pairing/clustering configurations, and the scope of inter- and intra- domain contacts within each of them. 

We hope that the present work will inspire systematic experimental investigation of looping models in providing a robust physical mechanism of long-range gene regulation. Collaborative efforts on this front are already in progress.

\section{Materials}
\subsection{Coarse-grained model of chromatin} 
\label{sec:methods} 
We use a coarse-grained model of chromatin for our simulations; a bead-spring polymer where the non-overlapping spherical beads represent nucleosomes and the springs connecting them represent the linker DNA and capture the overall flexibility of the nucleosomal packing. The spring is modeled by a version of the FENE potential~\cite{Binder1995}, which allows it finite extensibility, capturing chromatin's stiffness to stretching. An energy cost to bending of the springs models the bending rigidity of chromatin. We also introduce phantom beads to ensure that the model polymer is non-self-crossing and all configurations preserve linking number. The polymer is also self-avoiding; we introduce a highly repulsive $r^{-12}$ potential ($r$ is the center-to-center distance of beads) when beads overlap, effectively rejecting any MC step that leads to overlap. 

The transient interaction between nucleosomes is modeled as follows. As mentioned in the main text, the beads can be in two discrete states, \emph{active} and \emph{inert}, where the \emph{active} state is of higher energy. Two \emph{active} beads which happen to be within a short-interaction-range of each other (in any polymer configuration that the system explores) can form a temporary bond. This bond's attractive potential is parametrized by the \emph{attraction} parameter. The potential barrier to being \emph{active} is parametrized by the \emph{availability} parameter. In our simulations, we have used both a square well and a Gaussian well potential with a cut-off radius of twice the bead radius. Both of these potentials are a few $k_BT$ in strength, where $k_B$ is the Boltzmann constant and $T$ is the temperature. The beads stochastically transition between the two states independent of the chromatin-polymer dynamics. We consider a range of values of \emph{attraction} and \emph{availability} parameter in our simulations. Both of them control effective inter-bead interactions and we have only presented robust features from many such parameter choices. We also disallow a nucleosome to make multiple bonds with other nucleosomes, a condition we refer to as \emph{bond saturation}. Therefore, the higher-order interactions are repulsive in our model thereby disfavoring polymer collapse. It is known that the electrostatic-charge distribution on nucleosomes is highly complex~\cite{Arya2006}, and uniformly attractive higher order interactions seems unfeasible. 
		
We simulate our model using Monte Carlo (MC) method according to the standard Metropolis algorithm~\cite{Binder1995}. In order to save computational cost, the slower processes of update of the state of the nucleosomes and the status of the bonds between them are done asynchronously with exponential waiting times between queries, where a single bead is chosen randomly at random query times to attempt the update steps. All beads are taken to be identical, and the raw data is the set of temporary bonds formed between beads at fixed sampling time intervals for the entire run. The bead-spring polymer is in a ring construction with $200$ beads. We use a closed circle instead of linear molecule for two reasons.  The first is to avoid potential boundary effects at the ends of the polymer. The second is to model a topologically independent chromatin loop domain.   After an initial equilibration time, the equilibrated random configuration is used as the initial configuration for parallel runs to collect our data, which is of the order of fifty million accepted MC steps for \emph{each} bead (where each MC attempted step length is five percent of bead size).

\begin{acknowledgments}
We thank Pankaj Mehta for discussions and for a careful reading of the manuscript. This work was partially supported by NHGRI grant R01HG003470 to AMS.
\end{acknowledgments}


%
%

\end{document}